**Machine-learning-based head impact subtyping based on the spectral densities of the measurable head kinematics.**

Xianghao Zhan[1], Yiheng Li[2], Yuzhe Liu[1], Nicholas J. Cecchi[1], Samuel J. Raymond[1], Zhou Zhou[1], Hossein Vahid Alizadeh[1], Jesse Ruan[3], Saeed Barbat[3], Stephen Tiernan[4], Olivier Gevaert[2], Michael M. Zeineh[5], Gerald A. Grant[6], David B. Camarillo[1]

1. Department of Bioengineering, Stanford University, Stanford, CA, 94305, USA.

2. Department of Biomedical Data Science, Stanford University, Stanford, CA, 94305, USA.

3. Ford Motor Company, 3001 Miller Rd, Dearborn, MI 48120, USA.

4. Technological University Dublin, Dublin, Ireland.

5. Department of Radiology, Stanford University, Stanford, CA, 94305, USA.

6. Department of Neurosurgery, Stanford University, Stanford, CA, 94305, USA.

Corresponding author: Yuzhe Liu (yuzheliu@stanford.edu)

Xianghao Zhan and Yiheng Li contributed equally to this work.

**Abstract**

Objective: Traumatic brain injury can be caused by head impacts, but many brain injury risk estimation models are not equally accurate across the variety of impacts that patients may undergo and the characteristics of different types of impacts are not well studied. We investigated the spectral characteristics of different head impact types with kinematics classification.

Methods: Data was analyzed from 3,262 head impacts from lab reconstruction, American football, mixed martial arts, and publicly available car crash data. A random forest classifier with spectral densities of linear acceleration and angular velocity was built to classify head impact types (e.g., football, car crash, mixed martial arts). To test the classifier robustness, another 271 lab-reconstructed impacts were obtained from 5 other instrumented mouthguards. Finally, with the classifier, type-specific, nearest-neighbor regression models were built for brain strain.

Results: The classifier reached a median accuracy of 96% over 1,000 random partitions of training and test sets. The most important features in the classification included both low-frequency and high-frequency features, both linear acceleration features and angular velocity features. Different head impact types had different distributions of spectral densities in low-frequency and high-frequency ranges (e.g., the spectral densities of MMA impacts were higher in high-frequency range than in the low-frequency range). The type-specific regression showed a generally higher $R^2$-value than baseline models without classification.

Conclusion: The machine-learning-based classifier enables a better understanding of the impact kinematics spectral density in different sports, and it can be applied to evaluate the quality of impact-simulation systems and on-field data augmentation.

**Introduction**

Traumatic brain injury (TBI) is a growing public health hazard with high mortality and morbidity, as well as a socio-economic issue causing enormous diagnosis and treatment expenses[1]. This is particularly urgent for mild TBI (mTBI), given that mTBI is notoriously underreported, difficult to diagnose, and pose a potential predisposing factor to long-term neurodegenerative processes[2-4]. TBI/mTBI can be caused by various types of head impacts such as accidental falls, bike accidents, car crashes, American football impacts, mixed martial arts (MMA) impacts, water polo, ice hockey and car crashes[5-10]. Here, the different types of head impacts are defined as the different sources of impact (e.g., different contact sports).

Considering the consequences and prevalence of TBI/mTBI, various biomechanical studies have focused on the estimation of brain injury risk[11-16]. Physiologically, the brain is thought to be damaged by the inertial movement of the brain after the head sustains a physical impact or the rapid acceleration or rapid deceleration. Therefore, metrics of brain deformation are effective biomechanics predictors to predict TBI/mTBI. As a quantifier of the brain deformation, brain strain is generally recognized as a TBI/mTBI injury risk metric[12-15]. To calculate brain strain, the head kinematics, which can be measured with wearable accelerometers and gyroscopes, are the necessary input variables. However, the state-of-the-art approach to computing brain strain, the finite element modeling based on the brain physics, is computationally costly in terms of time and complex computational software. It typically takes hours to model the brain strain for one impact, which make it hard to be used in real-time monitoring of brain injury risks[12-14]. Therefore, researchers have developed many mathematical models (often referred to as the brain injury criteria) by reduced-order brain physics approximation and statistical fitting to rapidly estimate the brain injury risk from the head kinematics[9,11,17].

However, a recent study[18] found that different head impact types tend to have variable biomechanical characteristics, and the impact types should not be ignored when estimating the risk of TBI/mTBI. However, the brain injury criteria were developed based on certain types of head impacts[11,17], and therefore, should not be generally used across head impact types and the different kinematic features these brain injury criteria use can weigh differently across impact types[18,19]. To better develop risk evaluation models adaptable to various head impact types for detection and monitoring of TBI/mTBI, it is worthwhile to investigate the difference in the kinematics of various types of head impacts. Sports-specific monitoring and protection strategies can be developed if we understand the difference among types of head impacts.

To study the difference across head impact types, we used the kinematics of 3,262 head impacts from head model simulations, American football, MMA, automobile crashworthiness tests and car racing. We extracted the spectral densities of linear acceleration and angular velocity, classified these impacts with machine learning models, and then analyzed the most important features for classification. Finally, we used the classification model to build type-specific regression models of 95% maximum principal strain (MPS95), 95% maximum principal strain in corpus callosum (MPSCC95) and cumulative strain damage (CSDM, 15%, indicating the volume fraction of brain with MPS exceeding the threshold of 0.15[20]), and compared with a baseline model developed with a mixture of different types of head impacts. There metrics are chosen because previous studies have found correlation between these tissue-level biomechanics metrics and TBI[21-24].

**Materials and Methods**

**1. Data description**

To study a broad range of head impact types, we collected kinematics from a total of 3,262 head impacts from various sources: 2,130 laboratory head impacts (HM: head model) simulated from a validated finite element (FE) model of the Hybrid III anthropomorphic test dummy headform[14,25], 302 college football (CF) head impacts measured by the Stanford instrumented mouthguard[15,26], 457 MMA head impacts (MMA) measured by the Stanford instrumented mouthguard[10,27], 53 reconstructed head impacts with helmet from the National Football League (NFL)[28], 48 head impacts in automobile crashworthiness tests from NHTSA (NHTSA)[29], and 272 reconstructed head impacts from National Association for Stock Car Auto Racing (NASCAR).

## 2. Feature Extraction

To classify different types of head impacts, we extracted the spectral density features of the impacts because we believe different head impact types have different spectral characteristics. The features were extracted from the linear acceleration and angular velocity (four channels: three spatial components and the magnitude (the time-varying resultant of the three spatial components); x: posterior-to-anterior, y: left-to-right, z: superior-to-inferior), because they are directly measured by accelerometers. (Example impact kinematics was shown in Fig. 1).

Fast Fourier Transform (FFT) was applied to each channel of the kinematics, and the spectrum was split into windows, each with a width of 50Hz. We kept the first four windows because the four windows show high classification accuracy and frequency higher than 200Hz is viewed as noises in previous studies[14,15]. In each frequency window, the mean, maximum and median of the spectral density were extracted as the features. A total of 96 features (2 kinematics, 4 channels, 4 spectrum windows, 3 statistics) were extracted for each impact. (Feature heatmap was shown in Fig. 1). It should be mentioned that the window width was chosen to be 50Hz to enable at least 3 frequency points within each time window. The spectral feature extraction was performed with MATLAB R2021a (MathWorks, Natick, MA, USA).

### 3. Classification Algorithm and Evaluation

Over the past decades, there has been a rapid development of machine learning technology and machine learning technology has been used in the management of sport injury, the comprehension of sport behavior and the improvement of athlete performance[12,14,30-31]. In this study, to investigate the categorization of different types of head impacts, we applied the random forest as the machine learning classification algorithm to classify various types of head impacts. Random forest is a tree-based ensemble learning algorithm that builds multiple decision trees to classify the samples into different leaves via the minimization of Gini index or entropy[32-34]. Random forest builds trees with sub-samples of the dataset, adopts bootstrap aggregating (bagging), and performs a majority vote on the output of the trees. The reason to use random forest was that it does not suffer from overfitting based on bagging. It can also show the feature importance while not suffering from feature collinearity, which otherwise makes other interpretable classifiers (e.g., logistic regression) harder to interpret feature importance. The random forest was implemented with the Python package scikit-learn (version 0.24.1)[35]. For classification, the inputs are the 96 spectral features of the head kinematics, and the model outputs are the types of head impacts: HM/CF/NMMA/NFL/NHTSA/NASCAR.

To validate the feasibility of classifying different types of head impacts, we randomly partitioned the entire dataset of 3,262 impacts into 80% training set and 20% test set with stratified sampling over 1,000 repeats (1,000 experiments with different random seeds in the training/test set partitions to enable randomness in the modeling process and test model robustness). The hyperparameters of the classifier (the number of decision trees and the maximum depth of each tree) were tuned in a five-fold cross validation on the training set by optimizing the classification accuracy. The test data are used to evaluate the classification model performance and they are used after the models have been finalized.

The classification problem in this study follows a multi-class classification protocol: an impact is classified into one of the six categories (HM/CF/MMA/NFL/NHTSA/NASCAR). To assess the classification performance, and to assess whether the classifier biased towards certain classes, the multi-class classification accuracy (percentage of correct predictions in all test samples, e.g., an MMA impact is predicted as an MMA impact) and three binary classification metrics were used: the mean precision, the mean recall, and the mean area under the receiver operating characteristic curve (AUROC) of the 20% test impacts. Even though this is a multi-class classification problem, these binary classification metrics were investigated and averaged to evaluate the model performance without biasing towards the majority class (the largest dataset HM). As the precision (e.g., correct MMA predictions divided by all predicted MMA impacts), recall (e.g., correct MMA predictions divided by all MMA impacts) and AUROC are binary classification metrics, we averaged the three metrics after calculating them on the respective classification of each type of head impact (e.g., MMA vs. non-MMA, CF vs. non-CF) to reflect the overall binary classification performance across all impact types.

### 4. Important Feature Analysis

As previous studies found significantly different performance of brain injury risk estimation models across head impact types, with the classification model, we can interpret the most important features for kinematics classification to find the different spectral characteristics intrinsic to different types of head impact kinematics. The importance of a feature is calculated by the normalized total reduction of the classification criterion (Gini index or entropy) brought by a feature[32,33]. To ensure the result robustness, we recorded the normalized feature importance in the modeling of random forest classifiers over the 1,000 repeats. In each repeat, the feature importance was calculated on the 80% training data. Finally, the mean feature importance was calculated and ranked. Finally, we did an additional validation of the features by picking up the top 5, 10 and 20 important features and modeling the random forest classifiers, with the same four metrics calculated.

## 5. Brain Strain Regression with Classification

Upon verifying the feasibility of kinematics classification, we built type-specific brain injury risk evaluation models with the classifier to demonstrate an application of the kinematics classification. Rather than build a risk evaluation model for the mixture of all different head impact types, we chose the type-specific model to address the previously observed hardship of estimating brain injury risks across different head impact types with one single model[18].

We used the four major datasets (HM, CF, MMA, NASCAR) with the most impacts, and performed a k-nearest neighbor (KNN) regression of 95% maximum principal strain (MPS95), 95% maximum principal strain on corpus callosum (MPSCC95) and cumulative strain damage (CSDM) on the kinematics after partitioning the dataset into 80% training data and 20% test data with 20 repeats (20 experiments with different random seeds in the dataset partition process to test the model robustness under randomness). The 20% test data was unseen in both the classification model and regression model training datasets. We used these three metrics because strained-based metrics that directly summarize the brain deformation have shown superior injury predictability[21-24]. KNN was used as it did not require strong distribution and model assumptions. In the regression, the k nearest training impacts of a test impact were found based on Euclidean distance. Then, the MPS95/MPSCC95/CSDM     prediction     for     the     test     impact     is     the     averaged

MPS95/MPSCC95/CSDM of the k nearest training impacts. The hyperparameter k was tuned via a five-fold cross-validation on the 80% training data while optimizing the root mean squared error (RMSE). To prevent any data leakage, the testing set has been held out until the final evaluation stage of the type-specific regression strategy. Here, besides the spectral densities, we included the time-peaks of the linear acceleration and angular velocity (four channels for each). For one thing, we would select the kinematics which are directly measurable by sensors. For another, the time-peaks of the angular velocity have been shown to correlate well with MPS95 and are incorporated in the designs of many brain injury criteria[36,37]. The ground-truth MPS95/MPSCC95/CSDM values were given by the KTH model, which is a validated FE model[38].

Different from the classification models, the inputs of the regression models are the 104 kinematics features (including both the 96 spectral features used for head impact type classification and the 8 temporal features used for the regression) and the outputs are the MPS95/MPSCC95/CSDM. Classification results are used in the regression to classify a particular impact into a head impact type for type-specific regression models.

The baseline regression accuracy was given by using the 80% training data to build a KNN model and the 20% test data to assess the model coefficient of determination ($R^2$). Different from the baseline model, the classification-regression model first built a classifier on the 80% training data and built KNN models for each type of head impact. In the testing stage, the impacts were classified into one of the types of head impacts in the training set and then the MPS95/MPSCC95/CSDM associated with the test impact was calculated by the type-specific KNN regression model. Because most impacts were from the dataset HM, directly calculating the RMSE and $R^2$ would have led to biased estimates of regression accuracy. Therefore, on the test impacts, we calculated the RMSE and $R^2$ based on the ground-truth types of head impacts (HM/CF/MMA/NASCAR) and took an average over the four types to avoid the influence exerted by the majority dataset HM, since we want the model to be accurate across different types of head impacts. Finally, Wilcoxon signed-rank tests were done to test statistical significance on $R^2$ and RMSE as the Shapiro-wilk test rejected the data normality assumption.

## 6. Validation of the classifier on different instrumented mouthguards

To estimate the influence of instrumented mouthguard types on the classifier, we applied the classifier to 271 head impacts collected by five different mouthguards in the lab[39]: Stanford Instrumented Customized/Boiling-and-Bite, Prevent Customized/Boiling-and-Bite, SWA Customized. 54 impacts were analyzed for each mouthguard except 55 for SWA Customized mouthguards.

## Results

Firstly, we performed kinematics classification based on the 96 features with a random forest algorithm (model input: 96 spectral features of the kinematics, model output: type of the head impacts). The accuracy, mean precision, mean recall, and mean AUROC were shown in Fig. 2 A-D. The medians of 1) classification accuracy, 2) mean precision, 3) mean recall, and 4) mean AUROC were above 0.95, 0.93, 0.85, 0.92, respectively, which demonstrates the feasibility of classifying different types of head impacts. (Example confusion matrices showing the correct and wrong predictions were visualized in Fig. 3.)

Based on the classifier, according to the Method Section 4, the top 20/10/5 most important features were extracted over the 1,000 repeats of random dataset partitions. The features and their definitions are listed in Table 1. The 20 most important features included both angular velocity features and linear acceleration features. The different frequency ranges were found important in the classification: there were 6 features in the low-frequency range (0-50Hz) among the top 10 most important features, including the mean and median spectral density of the resultant angular velocity, the Y-axis angular velocity and the resultant linear acceleration. Among the other top-10 important features, there were 3 features in high the frequency range (150-200Hz) from the Y-axis and Z-axis linear acceleration. Additionally, among the top 20 features, there were 9 angular velocity features (7 from the magnitude and 2 from the spatial components) and 11 linear acceleration features (2 from the magnitude and 9 from the spatial components), which showed that for both measured kinematics, the magnitudes and the kinematic were informative components in the classification.

The distribution of the six datasets on the top 5 features is shown in Fig. 4 and the distribution on the other five features of the top 10 features is shown in Fig. 5. It was shown that on the top five features from the low-frequency range (0-50Hz), the MMA impacts had the lowest spectral densities, while NHTSA/HM/NFL impacts had higher spectral densities in this range, and the CF/NASCAR impacts generally had spectral densities higher than MMA impacts and lower than NHTSA/HM/NFL impacts. On the contrary, in the high-frequency range (100-200Hz) shown in Fig. 5, the MMA impacts had higher spectral densities while NHTSA/HM impacts had lower spectral densities shown in Fig. 4.

The classification performance on the 20/10/5 most important features was shown in Fig. 2 A-D: there was a general performance decline as the feature number decreased while the classifiers based on top 10 features still showed high classification performance with medians of 1) classification accuracy, 2) mean precision, 3) mean recall and 4) mean AUROC above 0.94, 0.88, 0.80, 0.90, respectively. These results showed the feasibility of the kinematics classification with the subsets of most important features.

Furthermore, to further validate the classifier's performance did not rely heavily on the type of instrumented mouthguard and did not overfit the specific mouthguards we used to collect the impact kinematics, we performed the classification of 271 lab impacts collected by different mouthguards and the results were shown in Fig. 2 E. All the impacts were classified into football-like types, and most of them were HM/NFL impacts, which used the same methodology to generate head impacts as these 271 impacts.

Finally, to test whether classification could improve brain injury risk estimation, we built the KNN regression models of MPS95/MPSCC95/CSDM with and without classification. The test $R^2$ averaged over four datasets is shown in Fig. 2 F-H (model input: 104 kinematics features, model output: MPS95/MPSCC95/CSDM). It was shown that the regression models with classification were significantly more accurate in MPSCC95 and CSDM regression ($p<0.05$) while similarly accurate in MPS95 regression ($p>0.1$). The results, in terms of RMSE, are reported in Table 2 where similar findings are shown: averaged across the four types of head impact impacts, the regression models with classification were significantly more accurate in MPSCC95 regression ($p<0.01$) and CSDM regression ($p<0.05$) while there was no statistical significance on MPS95 regression ($p>0.1$).

**Discussion**

In this study, we demonstrated that the machine learning classification model based on the spectral densities of the head impact kinematics showed high classification performance in categorizing different types of head impacts. With the classification, brain-strain metrics regression accuracy non-inferior to building a single model across impact types can be achieved. In this study, the MMA and college football impacts were measured by the Stanford instrumented mouthguard, while the head model simulated impacts and NHTSA impacts were both simulated with the Hybrid III anthropomorphic test dummy headform. Our additional validation on 271 lab-reconstructed impacts measured by 5 other mouthguards also showed most predictions were HM/NFL impacts which were also football-like impacts simulated/reconstructed with dummy heads. The results showed that the model generally successfully classified different types of head impacts. For the football-like impacts, the classifier can categorize them measured with different types of instrumented mouthguards.

As for the research contributions, firstly, the analysis of the most important features in the classification enables better understanding of the differences among head impact types. For instance, the NHTSA impacts have higher spectral densities in low-frequencies and lower spectral densities in high-frequencies, while the MMA impacts have lower spectral densities in low-frequencies and higher spectral densities in high-frequencies. Via the classification algorithm, we can investigate the key features that may determine the impact types and the sports and visualize the distribution of the spectral densities. In our previous study, we have found that different kinematics factors (e.g., angular velocity, angular acceleration) have different predictive power of the brain strain across the variety of head impacts[19]. For example, angular velocity features tend to be more predictive in MMA impacts while angular acceleration tend to be more predictive in football impacts. It has been shown by other researchers for short-duration impacts, the peak resultant angular velocity is better correlated with brain strain, while for long-duration impacts, the peak resultant angular acceleration is better correlated with brain strain[40]. Even though the definition of long/short duration for an impact is not defined for on-field impacts, in this study, after analyzing the frequency components of different types of features, the MMA impacts show more high-frequency component and thus closer to be a short-duration impacts. This fact may be able to explain the previous observation that the angular velocity features better predict brain strain for MMA impacts[19].

Secondly, we built classifiers for different types of head impacts and made the model trained on the entire dataset publicly available. As the previous study have shown the issues of generalizability of brain strain estimation models across different head impact types[18], we have shown that the classifiers can benefit the development of type-specific brain injury risk estimation models, which shows higher accuracy in brain strain regression in this study. As the classification is based on noisy patterns defined by humans (i.e., sports), this categorization may not capture the intrinsic types of sports. However, this noisy categorization of patterns works in the improvement of risk estimation accuracy. For example, for a new impact needing to be evaluated, even though it is measured from American football event, it may be classified into NASCAR considering its spectral density fingerprint and the overall performance of the classification-regression leads to an improvement in the accuracy.

Thirdly, as data from laboratory impacts are relatively easier to obtain than on-field data, such as MMA impacts, researchers can conduct domain adaptation in the future to generate more simulated on-field impacts with model deep learning techniques, such as generative adversarial network (GAN) for data augmentation. This classifier with high performance can be useful as the discriminator for the evaluation of the simulated impacts.

Furthermore, as is shown in our validation experiments across different mouthguards, the classifier successfully distinguished the lab-reconstructed football-like impacts from on-field college football impacts, which indicates that the football-like impacts generated on the dummy head with pneumatic impactor still cannot fully capture on-field college football characteristics.

Therefore, this classifier can be applied to evaluate the quality of dummy head impact reconstruction/simulation systems.

One additional potential application of this study is that the kinematics-classification-based type-specific regression of the strain-based metrics (MPS95/MPSCC95/CSDM) may help researchers rapidly estimate the strain-based metrics. It has been shown in previous studies that these strain-based metrics are good predictors of mTBI and associated pathologies (summary of these research can be found in the review[41]): for example, Wu et al. found 50% concussion thresholds of 0.270 for MPS95 and 0.477 for CSDM with human data[24]; Hajiaghememar et al. found a 50% axonal injury threshold of 0.286 for MPS95 in a large animal model[23]. To obtain these strain-based metrics, conventional state-of-the-art finite element models can take 7-8 hours for simulation per impact[14] (e.g., using a 16 GB RAM, Intel Core i7-6800 K CPU). However, with the rapid estimator discussed in this study, the computational time per impact can be dramatically reduced to be within 10 seconds (e.g., using an 8GB RAM, Intel Core i5-6300 U CPU). Therefore, the classification-based strain-metrics regression models can also be further applied in the field of TBI research.

As for the study limitations, first, to test our classifier does not rely heavily on the types of instrumented mouthguards, we only used football-like impacts measured by five mouthguards. In the future, more MMA impacts, NHTSA impacts measured by different devices can be collected and used to test the model's sensitivity to measurement devices on impacts other than football impacts. Second, to enable the classifier to be more accurate and broader in applications, more data from diverse types of head impacts should be collected and modeled. Additionally, we used the KTH model as the validated model to calculate brain strain. It is limited when compared to the recently developed state-of-the-art finite element head models (FEHM)[42-44]. For example, the KTH model does not model the gyri or sulci which have been shown to have significant influences on the FEHM behavior. In the future, more recently developed FEHMs can be applied to validate the results on the brain strain.

Another limitation of this study is that due to the hardship of head impact data collection in the real-world scenarios, the quantity and diversity of the datasets used in this study is limited. In the future, more classes of data for the head impact kinematics classification will be collected including rugby impacts, water polo impacts, and ice hockey impacts[8,45,46]. Additionally, the current impact datasets will also be enlarged to be more representative of the distribution for each type of head impacts, including different player positions within a single sport (e.g., line vs skill positions in American football)[47].

Finally, in this study, we obey the definitions of the classes in accordance with the human definitions of impact types based on the sources of impacts, such as college football impacts, mixed martial arts impacts, car crash impacts, etc. In future research, an alternative classification could be soft-labeling head impacts, i.e., an unknown impact can be described as probabilities of each head impact type. For example: An impact could be labeled as 10% MMA, 40% HM and 50% CF. Then, the weights could be used to investigate the characteristics of the unknown head impacts. Based on this, a potential approach for the regression between head kinematics and brain strain is to weight the regression output given by the type-specific regression models according to the probability of soft-labeling classification. The difficulty of this regression is that the conditional probabilities of head impact types given the kinematics should not be directly transferred to brain strain because of the high nonlinearity of the brain-skull system. According to previous exploration of the brain-skull system[11,19], we believe that the conditional probabilities of head impact type given the kinematics should be amended to the weights on brain strain. Additionally, kinematics clustering[48] which finds the impact clusters by breaking the human-defined impact types may also be able to find impact partitions according to the characteristics intrinsic to the kinematics features in a data driven manner. These impact partitions may better fit the data and potentially lead to further improvement of type-specific modeling of brain strain.

## Conclusion

In this study we performed the classification of different types of head impacts and demonstrated the feasibility of classification with high accuracy based on the spectral density of measurable head kinematics (i.e., linear acceleration and angular velocity). The important features for head impact classification included both low-frequency and high-frequency ranges, both linear acceleration and angular velocity. The classifier was also validated on 5 other instrumented mouthguards to test the model performance across different types of mouthguard measurement devices. Finally, this study exhibited non-inferior accuracy in the regression of brain strain with classification of different types of head impacts, rather than a single model for the mixture of all types of impacts together. The classification also reveals the difference of different types of head impacts in the frequency domain. The classifiers are publicly available for researchers to build better type-specific estimation models for brain injury risk.

## Acknowledgement

This research was supported by the Pac-12 Conference's Student-Athlete Health and Well-Being Initiative, the National Institutes of Health (R24NS098518) and Stanford Department of Bioengineering.

**Code and data availability**

The classification model, feature extraction code, example kinematics file and a user introduction are posted at: https://github.com/xzhan96-stf/kinematics_classifier.

**Declaration of interest**

The authors declare no conflict of interest.

(CSDM) Across Different Types of Head Impacts. Annals of Biomedical Engineering. 2022 Aug 3:1-2.

**Tables**

**Table 1. The ranking and definitions of the top 20 most important features in kinematics classification and the mean normalized importance values over 1,000 random dataset partitions.**

| Ranking | Meaning | Mean Normalized Importance |
|---|---|---|
| 1 | $\|\omega\|$: median spectral density in [0,50Hz] | 0.094 |
| 2 | $\omega_y$: median spectral density in [0,50Hz] | 0.062 |
| 3 | $\|\omega\|$: mean spectral density in [0,50Hz] | 0.056 |
| 4 | $\|a\|$: median spectral density in [0,50Hz] | 0.036 |
| 5 | $\omega_y$: mean spectral density in [0,50Hz] | 0.031 |
| 6 | $a_z$: max spectral density in [150, 200Hz] | 0.023 |
| 7 | $a_z$: max spectral density in [100, 150Hz] | 0.022 |
| 8 | $\|a\|$: mean spectral density in [0,50Hz] | 0.021 |
| 9 | $a_y$: max spectral density in [150, 200Hz] | 0.020 |
| 10 | $a_z$: mean spectral density in [150, 200Hz] | 0.019 |
| 11 | $\|\omega\|$: max spectral density in [0,50Hz] | 0.019 |
| 12 | $\|\omega\|$: median spectral density in [150,200Hz] | 0.019 |
| 13 | $\|\omega\|$: median spectral density in [50,100Hz] | 0.019 |
| 14 | $\|\omega\|$: mean spectral density in [50,100Hz] | 0.017 |
| 15 | $a_x$: mean spectral density in [0, 50Hz] | 0.016 |

| 16 | $a_z$:max spectral density in [150, 200Hz] | 0.016 |
| 17 | $|\omega|$:max spectral density in [50, 100Hz] | 0.015 |
| 18 | $a_z$:mean spectral density in [100, 150Hz] | 0.015 |
| 19 | $a_z$:max spectral density in [100, 150Hz] | 0.015 |
| 20 | $a_x$:median spectral density in [0, 50Hz] | 0.015 |

**Table 2. The mean root mean squared error (RMSE) of the k-nearest neighbor regression of MPS95, MPSCC95 and CSDM across different types of head impacts. HM: head model simulated impacts without helmet; CF: on-field college football impacts; MMA: on-field MMA impacts; NASCAR: NASCAR car crash impacts.**

|  |  | MPS95 | MPSCC95 | CSDM |
|---|---|---|---|---|
| HM | Baseline | 0.015 | 0.026 | 0.040 |
|  | Classification | 0.017 | 0.028 | 0.042 |
| CF | Baseline | 0.031 | 0.033 | 0.069 |
|  | Classification | 0.036 | 0.036 | 0.074 |
| MMA | Baseline | 0.036 | 0.045 | 0.072 |
|  | Classification | 0.034 | 0.040 | 0.062 |
| NASCAR | Baseline | 0.056 | 0.096 | 0.114 |
|  | Classification | 0.051 | 0.079 | 0.101 |
| Average | Baseline | 0.035 | 0.050 | 0.074 |
|  | Classification | **0.034** | **0.045** | **0.070** |

**Figures**

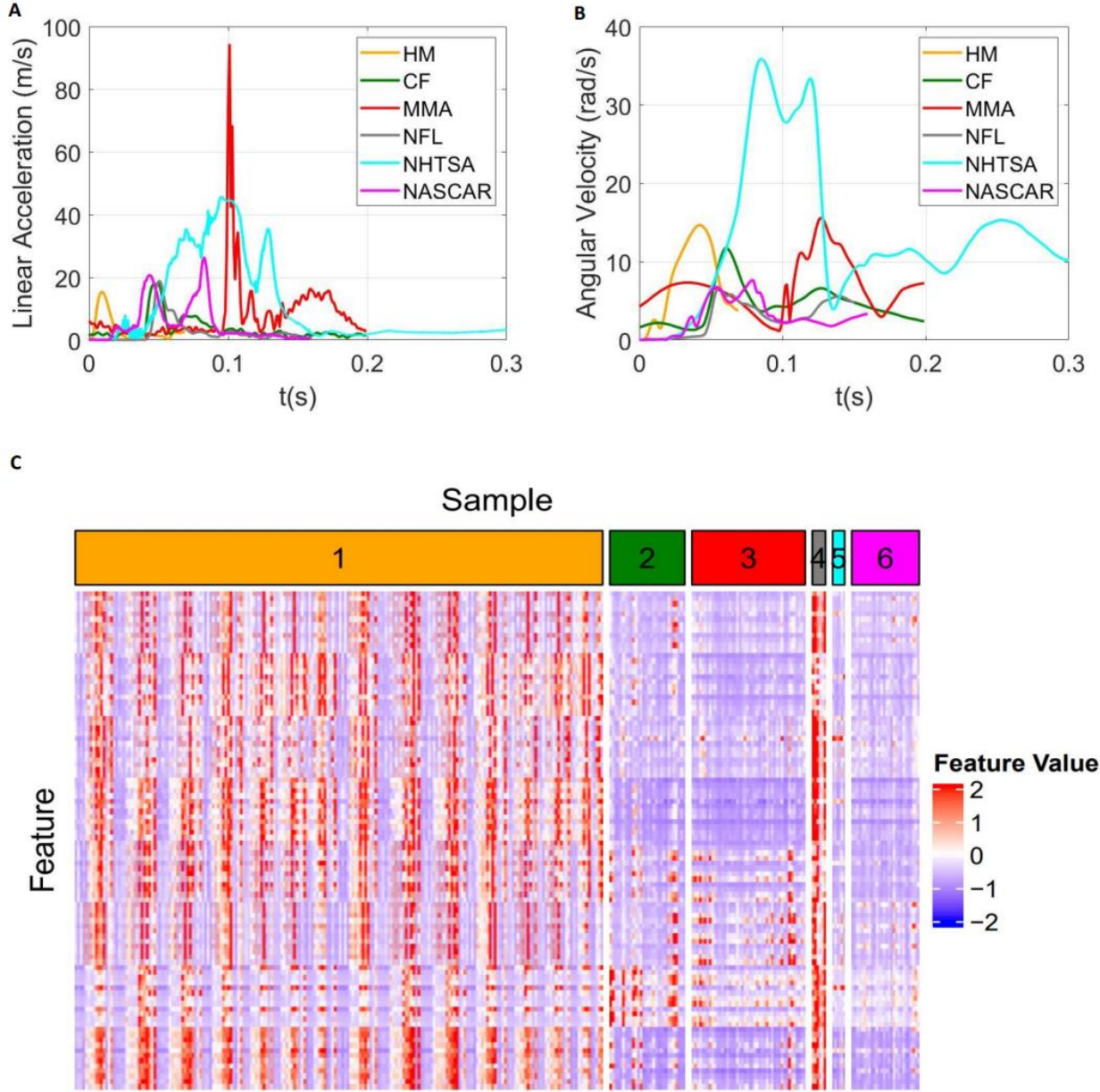

**Figure 1. Example kinematics of the six types of head impacts and the visualization of the six datasets used in this study with heatmap.** (A) The magnitude of linear acceleration at the brain center of gravity. (B) The magnitude of angular velocity. (C) The heatmap of features of all samples. 0-HM: head model simulated impacts without helmet, 1-CF: on-field college football impacts, 2-MMA: on-field MMA impacts,

3-NFL: lab-reconstructed NFL impacts with helmet, 4-NHTSA: NHTSA car crash impacts, 5-NASCAR: NASCAR car crash impacts.

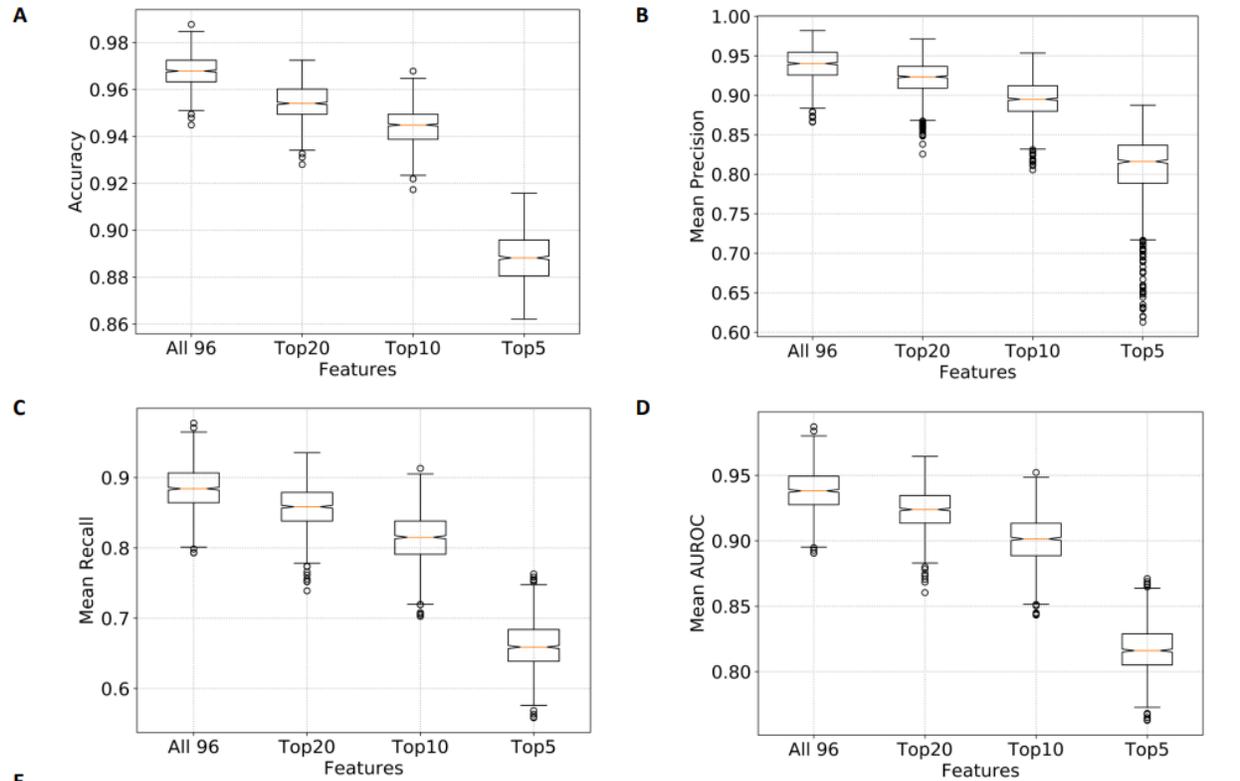

| | HM | CF | MMA | NFL | NHTSA | NASCAR |
|---|---|---|---|---|---|---|
| **Stanford Instrumented Boiling-and-Bite** | 39 | 2 | 0 | 13 | 0 | 0 |
| **Stanford Instrumented Customized** | 43 | 1 | 0 | 10 | 0 | 0 |
| **Prevent Boiling-and-Bite** | 44 | 0 | 0 | 11 | 0 | 0 |
| **Prevent Customized** | 50 | 0 | 0 | 4 | 0 | 0 |
| **SWA Customized** | 29 | 0 | 0 | 25 | 0 | 0 |

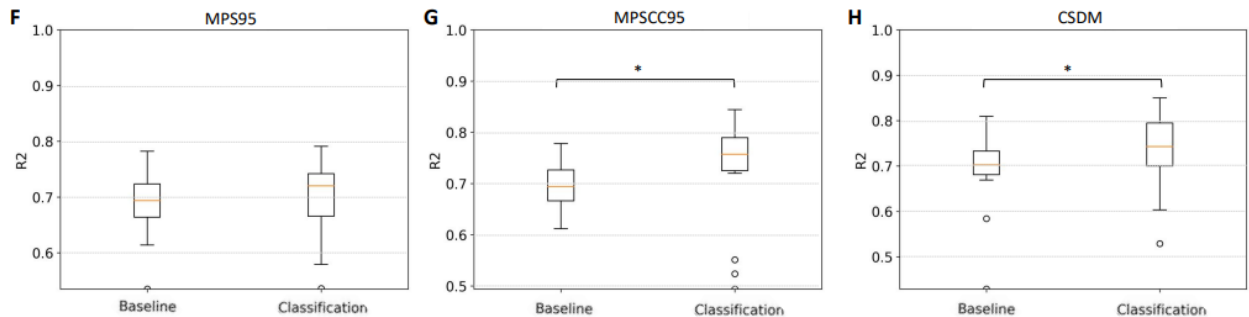

**Figure 2. Classification performance metrics of the random forest classifier and the MPS95/MPSCC95/CSDM regression accuracy with/without kinematics classification.** The accuracy (A), mean precision (B), mean recall (C) and mean AUROC (D) of the classification based on different numbers of features over 1,000 random dataset partitions. The prediction results of 271 lab-reconstructed football-like impacts measured by 5 different instrumented mouthguards (E). The mean regression $R^2$ of MPS95 (F), MPSCC95 (G) and, CSDM (H). 1000 random train-test partitions were done in the regression. (*:$p<0.05$, Wilcoxon signed-rank test)

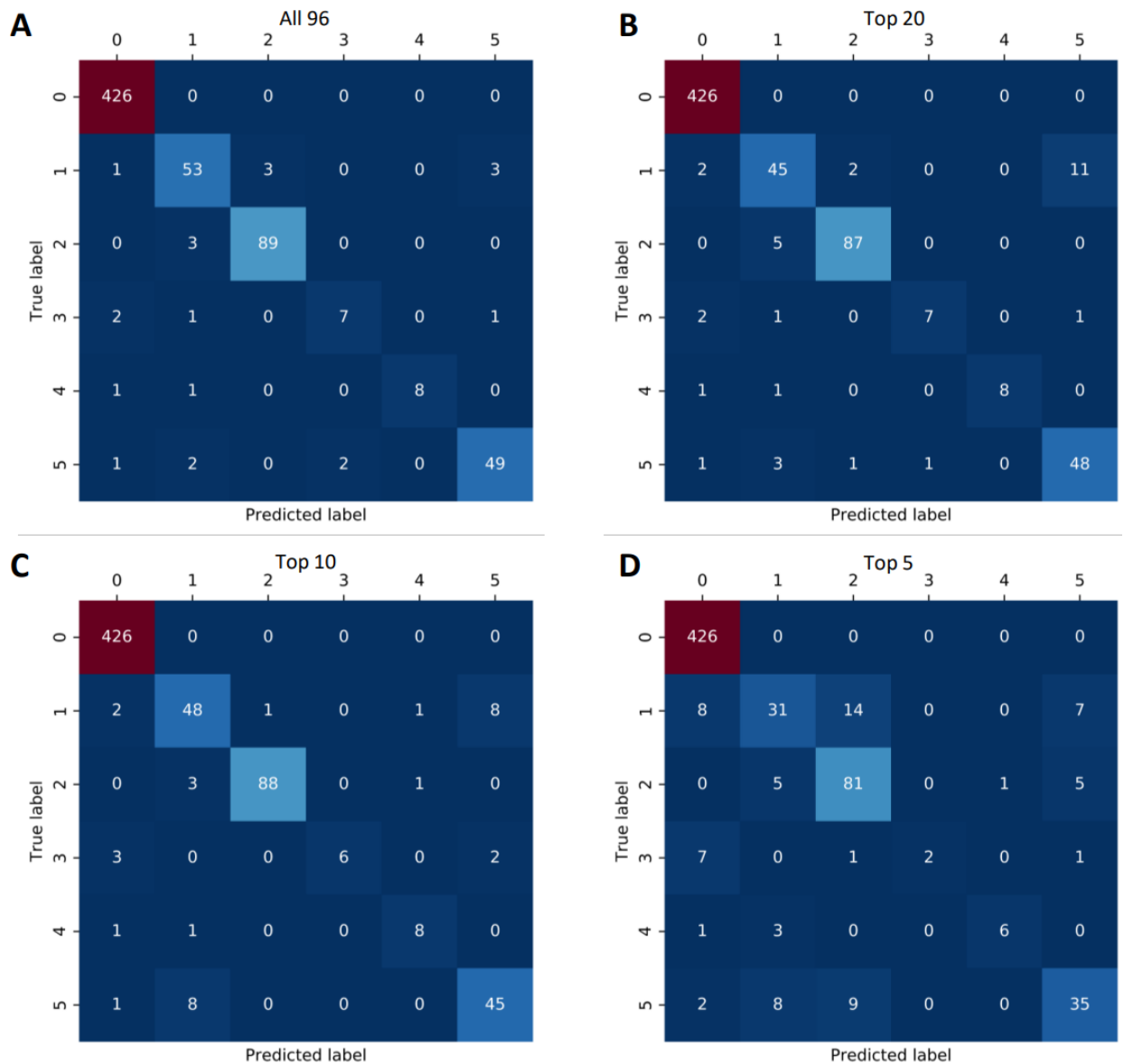

**Figure 3. The example confusion matrices of the classification based on four different numbers of features.** The confusion matrices for (A) all 96 features, (B) top 20 features, (C) top 10 features, and (D) top 5 features. 0-HM: head model simulated impacts without helmet, 1-CF: on-field college football impacts, 2-MMA: on-field MMA impacts, 3-NFL: lab-reconstructed NFL impacts with helmet, 4-NHTSA: NHTSA car crash impacts, 5-NASCAR: NASCAR car crash impacts.

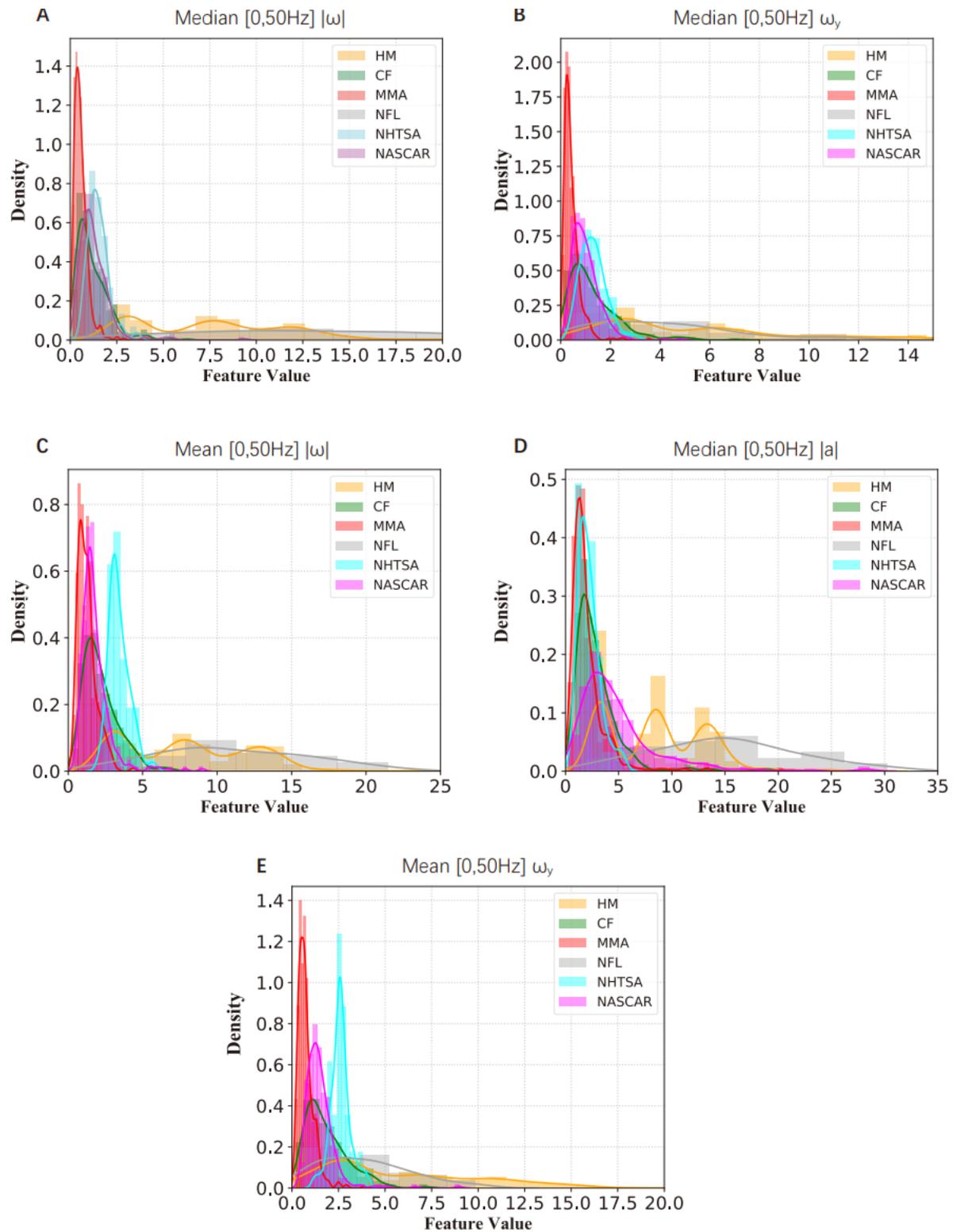

**Figure 4. The distribution of the six datasets on the top 5 most important features for classification.**

The data distribution in the median spectral density in [0,50Hz] of the resultant angular velocity (A), the median spectral density in [0,50Hz] of the Y-axis angular velocity (B), the mean spectral density in [0,50Hz] of the resultant angular velocity (C), the median spectral density in [0,50Hz] of the resultant linear acceleration (D), and the mean spectral density in [0,50Hz] of the Y-axis angular velocity (E). HM: head model simulated impacts without helmet, CF: on-field college football impacts, MMA: on-field MMA impacts, NFL: lab-reconstructed NFL impacts with helmet, NHTSA: NHTSA car crash impacts, NASCAR: NASCAR car crash impacts.

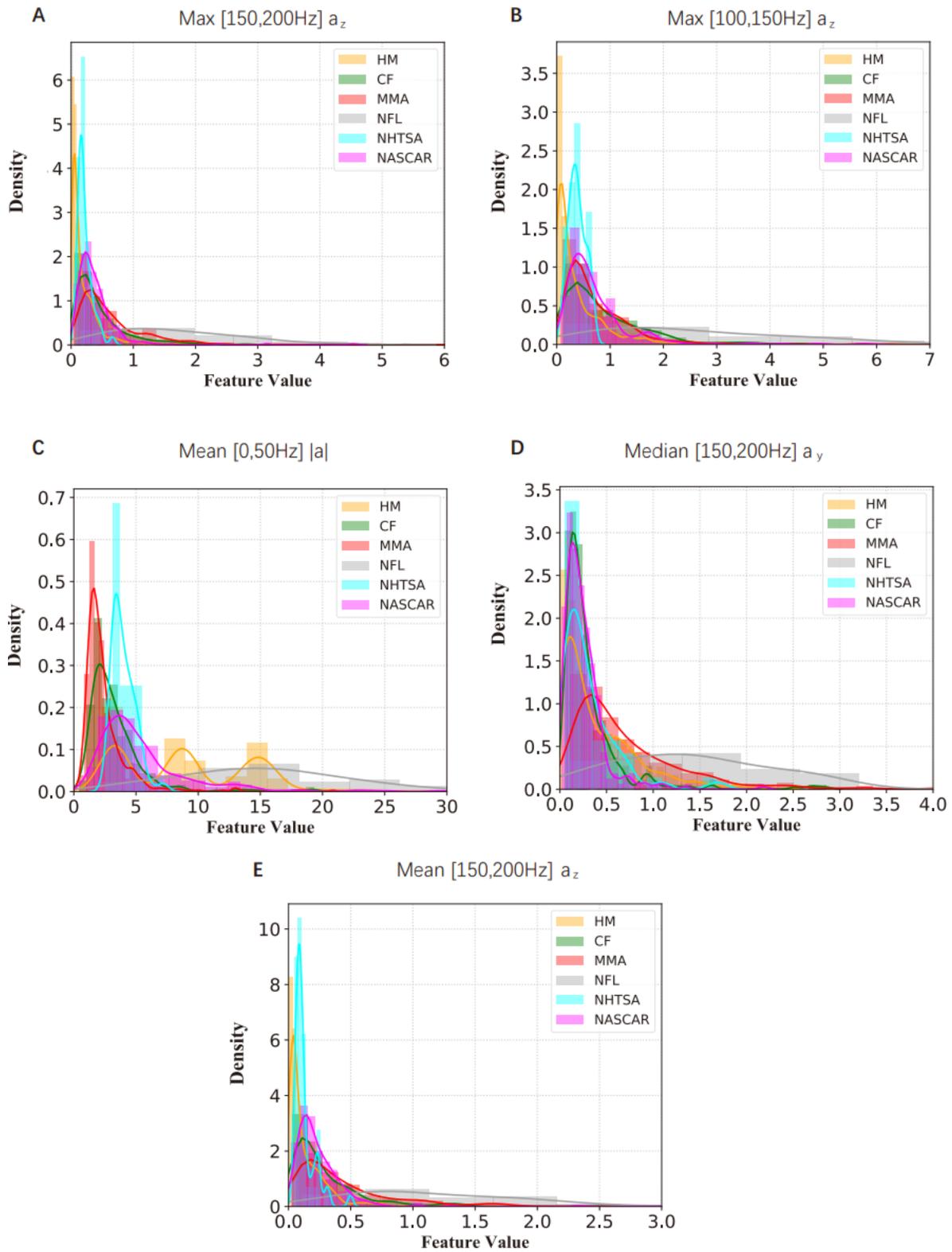

**Figure 5. The distribution of six datasets on the sixth-to-tenth most important features for classification.** The data distribution in the max spectral density in [150, 200Hz] of the Z-axis linear

acceleration (A), the max spectral density in [100, 150Hz] of the Z-axis linear acceleration (B), the mean spectral density in [0,50Hz] of the resultant linear acceleration (C), the max spectral density in [150, 200Hz] of the Y-axis linear acceleration (D), the mean spectral density in [150, 200Hz] of the Z-axis linear acceleration (E). HM: head model simulated impacts without helmet, CF: on-field college football impacts, MMA: on-field MMA impacts, NFL: lab-reconstructed NFL impacts with helmet, NHTSA: NHTSA car crash impacts, NASCAR: NASCAR car crash impacts.